\documentclass[10pt,aps,groupedaddress,twocolumn]{revtex4}
\usepackage{graphicx,amsmath,amsfonts}
\begin{document}
\newcommand{\be}{\begin{equation}}
\newcommand{\ee}{\end{equation}}
\title{Kondo effect in d-wave superconductors}
\author{Anatoli Polkovnikov}
\email{anatoli.polkovnikov@yale.edu}
\homepage{http://pantheon.yale.edu/~asp28}
\affiliation{Department of Physics, Yale University\\
P.O. Box 208120, New Haven, CT 06520-8120, USA}

\begin{abstract}
\vspace{0.5 cm} We present theoretical investigation of a single
magnetic impurity in a d-wave superconductor using the large N
limit. It is shown that the Kondo screening occurs only in the
presence of the particle-hole asymmetry. We find analytical
expressions for the Kondo temperature, magnetic susceptibility and
scattering matrix near the phase transition. The results are
generalized for the density of states vanishing with an arbitrary
exponent. Also we briefly study the modifications of the theory
for the case of a non-magnetic impurity which induces a staggered
spin configuration on the nearby copper atoms.
\end{abstract}

\pacs{7462.Dh, 7215.Qm,7472.-h,7520.Hr,7530.Cr}

\maketitle

\section{Introduction}
Recent experiments in d-wave BSCCO superconductors showed some
interesting phenomena in the presence of impurities on copper
sites. In particular, STM showed a strong variation of the
electron density of states (DOS) near Ni~\cite{Hudson,Hudson1} and
Zn~\cite{Davis} atoms in Bi$_2$Sr$_2$CaCu$_2$0$_{8+\delta}$. Using
NMR Kondo-like effect was in cuprates doped by Zn, Ni or Li
impurities both above and below
$T_c$~\cite{Alloul,Farlane,Mahajan,Bobroff,Mendels}, and an
uncompensated spin $S=1/2$ appeared near the Zn
impurity~\cite{Julien}. Although it is tempting to explain the
tunneling experiments by the pure potential scattering~\cite{Bal},
there are several key features which are not easily justified
within this model (see e.g.~\cite{Flatte}). For example, the low
frequency resonance can be achieved only for a large non-universal
value of the potential coupling~\cite{Atkinson}. On the other
hand, the Kondo screening gives a natural low energy scale and the
results appear to be much more robust with respect to the
variation of the composition and coupling constants~\cite{polk}.
We note that using STM data alone it is not possible to
distinguish between the potential and Kondo scattering mechanisms
leading to the variations of the DOS . The screened impurity acts
as a spin singlet and is therefore similar to the potential
scatterrer. However the unitarity limit in the Kondo case occurs
at the coupling slightly exceeding the critical value, which is
not necessarily large. The other important advantage of the Kondo
mechanism is that it simultaneously explains NMR and STM
measurements.

The cuprates are unconventional superconductors with linearly
vanishing density of states near the four nodes. The usual Kondo
picture is not valid
there~\cite{Withoff,Ing1,Fradkin,Fradkin1,Zhu,Zhang}. In
particular, there is no phase transition if the spin-spin coupling
is relatively small so the impurity remains effectively decoupled
from quasiparticles. If the DOS vanishes at the Fermi surface
slower than the square root of energy: $\rho(\varepsilon)\propto
|\varepsilon|^r$  with $r<1/2$, then it is possible to develop a
perturbation theory~\cite{Withoff}, which predicts Kondo
transition  at the spin-spin coupling larger than a certain
critical value proportional to $r$. However if $r\geq 1/2$ this
simple picture doesn't work. Thus both numerical renormalization
group (NRG)~\cite{Ing2,Vojta1} and dynamic large N multichannel
approach~\cite{Vojta} showed that the Kondo phase exists only if
the Hamiltonian doesn't respect the particle-hole symmetry. In the
symmetric case the spin-spin coupling always flows to zero. On the
other hand, in~\cite{Chen} it was argued that if the particle-hole
symmetry is preserved, then there is no screening at the level of
the total impurity susceptibility $T\chi_{imp}\to 1/4$ at $T\to
0$, but still there is a quenching of the impurity spin and
$T\chi_{loc}\to 0$ in the zero temperature limit.

The model Hamiltonian for a pure superconductor to be exploited in
the present paper is:
\begin{equation}
H_0=\sum\limits_{\bf  k} \psi^\dagger_i ({\bf k})
\left(\varepsilon_{\bf k} \tau_z^{ij}+\Delta_{\bf
k}\tau_x^{ij}\right) \psi_j({\bf k}),
\label{ho}
\end{equation}
where $\varepsilon_{\bf
k}=-\varepsilon_0(\cos{k_x}+\cos{k_y})-\mu$ and $\Delta_{\bf
k}=\Delta_0(\cos{k_x}-\cos{k_y})$ are the kinetic energy and
superconducting gap respectively, $\psi(k)$=$(c_{\uparrow}(k),
c_\downarrow^\dagger (-k))$ is the Gor'kov-Nambu spinor. The
interaction with a single impurity can be expressed in the
Coqblin-Schrieffer form~\cite{Hewson}:
\begin{eqnarray}
&&H_{int}=U\psi^\dagger_i ({\bf r}_0)\tau_z^{ij}\psi_j ({\bf
r}_0)\nonumber\\&&-{J\over 2} (\psi_{\uparrow}^\dagger({\bf r}_0)
F_\uparrow-F_\downarrow^\dagger \psi_\downarrow({\bf r}_0))
(F_\uparrow^\dagger\psi_\uparrow({\bf
r}_0)-\psi_\downarrow^\dagger({\bf
r}_0)F_\downarrow),\hspace{0.7cm} \label{hint1}
\end{eqnarray}
where $U$ and $J$ are the potential and spin-spin coupling
constants, $F$ is the effective impurity spinor in the Nambu
representation (it is related to the creation and annihilation
operators by
$F_\uparrow=f_\uparrow,\;F_\downarrow=f_\downarrow^\dagger$). The
single impurity occupancy corresponds to the condition:
\begin{equation}
f_i^\dagger f_i=1 \quad\Leftrightarrow\quad F_i^\dagger
\tau_z^{i,j} F_j=0 . \label{49}
\end{equation}
In fact, there is an additional potential like scattering term
proportional to $J$ in (\ref{hint1}), however, it disappears in
the path-integral formulation of the problem~\cite{Fradkin}. If
$\mu=0$ then the Hamiltonian (\ref{ho}) respects the particle hole
symmetry, i.e. under the transformation $k_x\to k_x+\pi,\;k_y\to
k_y+\pi$ both the kinetic energy $\varepsilon_k$ and the order
parameter $\Delta_k$ change their signs. This additional symmetry
is not generic, it disappears if either chemical potential or
second neighbor hopping becomes not zero. However this notion is
very useful, since in many cases the particle-hole asymmetry can
be treated as a perturbation. The model above but without the
potential scattering term was already examined by C. Cassanello
and E. Fradkin in~\cite{Fradkin,Fradkin1}. The authors concluded
that there is a Kondo phase transition above some critical
spin-spin coupling $J_c$. They found that for the linear DOS, the
Kondo temperature vanishes exponentially near the phase
transition. Also it was argued that in the Kondo phase the
impurity susceptibility vanishes at $T\to 0$ as $T\ln{1/T}$.
However, these results were obtained even in the particle-hole
symmetric case. We find similar dependences of the Kondo
temperature and susceptibility only if the particle-hole symmetry
is broken, e.g. by non-zero potential scattering. The reason in
such a discrepancy is that the symmetrization of the action with
respect to positive and negative frequencies used
in~\cite{Fradkin, Fradkin1} is not valid, which resulted in
essential inconsistencies there. We come back to this point in the
next section. The other difference between our approaches is that
we start from the Hamiltonian (\ref{ho}) instead of its linearized
nodal version~\cite{Fradkin, Fradkin1}. This is important for the
analysis because (i) the particle-hole asymmetry ($U\neq 0$ or
$\mu\neq 0$) is essentially a high energy effect, (ii) the spatial
structure of the scattering can not be reproduced from the nodal
version of the Hamiltonian, (iii) the expressions for the
$T$-matrix and Green functions are much more apparent in the
original Gor'kov-Nambu representation than in that of the nodal
particles. Although at the later stage we use linearized version
of the Hamiltonian for the sake of simplicity and analyticity, the
results can be easily generalized for the arbitrary spectrum of
quasiparticles~\cite{polk}.

We also study in detail the phase diagram at finite temperature,
where the effects of particle hole asymmetry become crucial. The
model is generalized to the case of an arbitrary exponent $r$
larger than $1/2$, where the simple Withoff-Fradkin
picture~\cite{Withoff} is invalid. We would also mention that the
problem addressed here but for the Anderson magnetic impurity was
studied in~\cite{Zhu} using the slave-boson mean field approach.
The impurity potential had been chosen in such a way that only
zero or single occupancy was allowed. As a result the Hamiltonian
was strongly particle-hole asymmetric from the beginning. It was
found that this model predicts the Kondo effect and the particle
hole symmetry of the pure superconductor is not important. The
model also results in a low-energy resonance in the DOS and is
consistent with the experimental spatial dependence.

The paper is organized as follows. In Section II we solve the
saddle point equations and study the phase diagram in the space of
the potential and spin-spin couplings $U$ and $J$. It is found
that the transition to the Kondo phase occurs only in the
particle-hole asymmetric case. At zero temperature the critical
spin-spin coupling $J_c$ saturates as $U\to 0$ if $r\leq 1$ and
diverges as $1/\sqrt{U}$ if $r>1$. In the case of a d-wave
superconductor ($r=1$), the Kondo temperature is found to vanish
exponentially near $J=J_c$ in agreement with~\cite{Fradkin1}. At
the fixed temperature and for $r\leq 1$ the crossover from the
high temperature ($T\gg T_K$) to the low temperature ($T\ll T_K$)
limit occurs at $J$ diverging as $U\to 0$.

In Section III we examine the expressions for the local and total
magnetic susceptibilities of the impurity and the quasiparticle
density of states. It is shown that in the particle-hole symmetric
case for $r\leq 1$ and $J>J_c$ the local susceptibility diverges
as $T\to 0$, but $T\chi_{loc}\to 0$, while $T\chi_{imp}\to {\rm
const}$. In the Kondo phase $\chi_{loc}$ saturates and
$\chi_{imp}\to 0$. Both $\chi_{loc}$ and $\chi_{imp}$ as functions
of temperature have a maximum at $T\approx T_K$. In the end of the
section we derive the expression for the induced quasiparticle DOS
near the impurity. It is shown that in the Kondo phase the
effective scattering matrix has a pole at the frequency close to
the Kondo temperature ($\Omega\approx T_K$). The crucial
difference between this model and simple potential
scattering~\cite{Bal} is that the unitarity limit here occurs at
finite magnitude of the spin-spin coupling $J\approx J_c$, which
doesn't have to be large.

Section IV is devoted to the discussion of the modifications of
the model for the case of a non-magnetic impurity, which induces a
staggered magnetization nearby. This study is motivated by the NMR
experiments~\cite{Alloul,Julien} unambiguously showing existence
of the magnetic moments in the vicinity of the Zn atom, which
substitutes Cu. From the theoretical point of view this issue is
interesting, since the spin-spin interaction becomes non-local.
Assuming that the effective spin is induced on the first Cu
neighbors of Zn we show that the action has four different saddle
points, and for the whole range of coupling constants the
configuration with the $D$ symmetry of the order parameter
corresponds to the lowest energy. The critical spin-spin coupling
for the non-local spin  is found to decrease with $U$ as opposed
to the local spin case. The spatial dependence of the induced DOS
is also different. It is characteristic for the scattering on the
four rather than the single site and gives an excellent agreement
with the STM experiments~\cite{Davis}.

\section{Saddle point approximation for a Single magnetic impurity}

After the standard Hubbard Stratonovich decoupling and integrating
out Fermion fields the partition function for the system with the
Hamiltonian given by (\ref{ho}) and (\ref{hint1}) becomes a
functional integral over the auxiliary fields $\phi$ and
$\epsilon$ with the action:
\begin{eqnarray}
S=&-&{\rm Tr}\ln\left({\partial \over \partial
\tau}+\epsilon(\tau)\tau_z -\phi(\tau)
G\left({\partial\over\partial \tau}\right)
\phi(\tau)\right)\nonumber\\&+&{2\over J}\int\limits_0^\beta
\phi(\tau)^2d\tau, \label{1}
\end{eqnarray}
where trace is taken over all antiperiodic functions of~$\tau$
$(f(\tau+\beta)=-f(\tau))$; $G(\partial/\partial \tau)\equiv
G({\bf r}_0, {\bf r}_0,\partial/\partial\tau)$ is the
quasiparticle Green function in the presence of the potential
scattering on the impurity:
\begin{widetext}
\begin{equation}
G({\bf r},{\bf r}^\prime,\omega_n)=G^0({\bf r}-{\bf
r}^\prime,\omega_n)-G^0({\bf r}-{\bf r}_0,\omega_n)
U\tau_z\left(1+U\tau_z G^0(0,\omega_n)\right)^{-1}G^0({\bf r
}_0-{\bf r}^\prime,\omega_n),
\end{equation}
\end{widetext}
and $G^0$ is the free electron Green function, which in the
momentum space is equal to:
\begin{equation}
G^0_{\bf k}(\omega)=\left(-i\omega+\varepsilon_{\bf k}\tau_z+
\Delta_{\bf k}\tau_x\right)^{-1}.
\end{equation}
The Lagrange multiplier $\epsilon$ in (\ref{1}) enforces the
single occupancy constraint. In general $\phi$ is a complex field,
however its phase can be reabsorbed into $\epsilon$ by a simple
gauge transformation~\cite{Read}.

It is not hard to show that in a d-wave superconductor the free
particle Green function at ${\bf r}=0$ is expressed as~\cite{Bal}:
\begin{equation}
G^0(0,\omega)=G^0_0(\omega)+G^0_1(\omega)\tau_z
\label{2}
\end{equation}
with $G^0_0$ and $G^0_1$ being even and odd functions of
frequency, respectively. If the free Hamiltonian is also invariant
under the particle-hole transformation then $G^0_1(\omega)\equiv
0$. For simplicity we assume this is the case. In fact, it can be
shown that the models with $U\neq 0$ and $G^0_1\neq 0$ can be
mapped to each other. Explicitly $G^0_0(\omega_n)$ is given by
\begin{equation}
G^0_0(\omega_n)=\sum\limits_{\bf k} {i\omega_n\over \omega_n^2+
\varepsilon_{\bf k}^2+\Delta_{\bf k}^2}.
\end{equation}
At small frequencies the main contribution to $G^0_0$ comes from
the wavevectors close to the nodes, so we can do the summation
over ${\bf k}$ and obtain:
\begin{equation}
G^0_0(\omega_n)\approx{i\omega_n\over \pi v^2}
\ln{\omega_n^2+\Lambda^2\over \omega_n^2},
\label{G0}
\end{equation}
where $v=\sqrt{v_F v_\Delta}$ is the mean geometrical velocity
near the nodes and $\Lambda$ denoting the upper cutoff of the
order of $\Delta$. For the Hamiltonian (\ref{ho}) with $\mu=0$,
$v_F=\varepsilon_0\sqrt{2}$, $v_\Delta=\Delta_0\sqrt{2}$.

From the equation (\ref{2}) we see that the action (\ref{1})
splits into two identical terms for the spin up and spin down
polarizations so that:
\begin{eqnarray}
S=&-&N\,{\rm Tr}\, \ln\left({\partial\over\partial\tau}
+\epsilon-\phi^\dagger
(\tau)\,G_\uparrow\left({\partial\over\partial\tau}
\right)\phi(\tau)\right)\nonumber\\&+&{N\over
J}\int\limits_0^\beta d\tau |\phi(\tau)|^2-{N\over
2}\epsilon\beta,
\label{main}
\end{eqnarray}
where $N=2$ is the number of the spin channels,
\begin{equation}
G_\uparrow
\left({\partial\over\partial\tau}\right)=G^0_0\left({\partial\over\partial\tau}
\right)\left(1+UG^0_0\left({\partial\over\partial\tau}
\right)\right)^{-1}.
\label{a1}
\end{equation}
Note that there is an additional term "$-N/2\,\epsilon\beta$" in
(\ref{main}), which enforces the symmetry $\epsilon\to -\epsilon$
in the absence of the potential scattering. In (\ref{main}) we
find it more convenient to use a gauge, where $\phi$ is complex,
but $\epsilon$ is time independent. In the large $N$ limit the
action is given by the saddle point approximation~\cite{Read}. So
it is necessary to find the stationary point with respect to the
auxiliary fields $\epsilon$ and $\phi$. Let us start with
$\epsilon$:
\begin{equation}
{2\over N}{\partial S\over \partial \epsilon}\!\equiv
I(\epsilon)\!=\!-\beta\!-\!\!\sum_{\omega_n}{2\over {{\rm
e}^{-i\omega_n\!\delta}-1\over\delta}\!+\!\epsilon\!-\!
G_\uparrow(\omega_n)|\phi|^2},
\label{seffeps}
\end{equation}
where $\delta$ is an infinitesimal parameter showing the correct
procedure of closing the Wick's contour. If $U=0$ then for any
$\phi$:
\begin{equation}
I(\epsilon)=\beta(-1+2 g_F(\epsilon))\quad \Longleftrightarrow
\quad I(\epsilon)+I(-\epsilon)=0, \label{QQ}
\end{equation}
where $g_F$ is the Fermi function. In fact this identity is just a
consequence of the particle-hole symmetry. If $U\neq 0$ the
relation above generalizes to
\begin{equation}
I(\epsilon,U)+I(-\epsilon,-U)=0.
\end{equation}
So in the absence of the potential scattering, $\epsilon=0$ at the
saddle point. With help of (\ref{QQ}), we write the saddle point
condition as follows:
\begin{eqnarray}
&&\sum_{\omega_n} {-\epsilon\over
\left(i\omega_n\!-\!\epsilon\!+\!|\phi|^2
G_\uparrow(\omega_n)\right)
(i\omega_n+|\phi|^2G_0^0(\omega_n))}\nonumber\\=&&\sum_{\omega_n}
{U|\phi|^2 G_0^0(\omega_n)G_\uparrow(\omega_n) \over
\left(i\omega_n\!-\!\epsilon\!+\!|\phi|^2
G_\uparrow(\omega_n)\right) (i\omega_n\!+\!|\phi|^2
G_0^0(\omega_n))},\phantom{XX} \label{1u}
\end{eqnarray}
so that both sums become convergent at $\omega_n\to\infty$.
Clearly $\epsilon=0$ as long as $|\phi|=0$. Near the phase
transition the order parameter $|\phi|$ is small and therefore
$\epsilon$ is also small. After introducing dimensionless
parameters
\begin{equation}
\epsilon\!\to\! {\epsilon\over \Lambda},\;
\beta\!\to\! \Lambda\beta,\; U\!\to\! {\Lambda U\over \pi v^2},\;
J\!\to\! {\Lambda J\over \pi v^2},\; |\phi|^2\!\to\!
{|\phi|^2\over \pi v^2}
\end{equation}
and using $\epsilon\ll 1$ we can simplify (\ref{1u}) to:
\begin{equation}
\tanh{\epsilon\beta\over 1+2|\phi|^2\ln{1\over |\epsilon|}}
\approx \left(2+4|\phi|^2\ln{1\over |\epsilon|} \right)F(\phi,U),
\label{2u}
\end{equation}
where
\begin{equation}
F(\phi,U)=\int\limits_0^\infty {dx\over\pi} {U|\phi|^2
\over\left[\ln^{-1}\left(1+{1\over x^2}\right)+|\phi|^2\right]^2
+U^2x^2}. \label{eq:23}
\end{equation}
%
The function $F$ increases linearly with $U$ at small $U$ and
saturates at $1/2$ for large $U$. At finite temperature and small
$U$ or $|\phi|$, (\ref{2u}) shows that $\epsilon\propto T$ up to
logarithmic corrections, i.e. instead of (\ref{2u}) we can write:
\begin{equation}
\tanh{\varepsilon\beta\over 1+2|\phi|^2\ln{\beta}}\approx 2
\left(1+2|\phi|^2\ln{\beta}\right) F(\phi,U). \label{3u}
\end{equation}
As long as $T$ is not too small we can ignore weak logarithmic
dependence of the RHS of (\ref{3u}) on $T$, and $\epsilon$ is a
decreasing function of temperature. This picture is valid down to
\begin{equation}
T_K(\phi,U)\approx \exp\left(-{1-2F(\phi,U)\over 4|\phi|^2
F(\phi,U)}\right), \label{T_K}
\end{equation}
at this temperature $\epsilon$ saturates becoming
\begin{equation}
\epsilon_0\approx T_K \left(1+2|\phi|^2\ln{1\over T_K}\right).
\label{24}
\end{equation}
The quantity $T_K$ gives the low-energy scale of the problem and
further will be identified with the Kondo temperature. Note that
it has an exponential dependence on the order parameter $|\phi|$.

The second equation defining the saddle point is
obtained differentiating (\ref{main}) with respect to $\phi$.
There is always a trivial solution $\phi=0$, the second one
is found from:
\begin{equation}
{1\over J}=\int\limits_0^\infty {d\omega\over \pi} \Re\left\{
{G_\uparrow(\omega)\over i\omega-\epsilon +|\phi|^2
G_\uparrow(\omega)}\right\},
\label{seffphi}
\end{equation}
where the summation over discrete $\omega_n$ was transformed to
the integration, which is valid in the low temperature limit.
Assuming $\epsilon\ll 1$ and $\epsilon U\ll 1$, i.e. the system is
in the vicinity of the phase transition, (\ref{seffphi}) becomes:
\begin{equation}
{1\over J}\!=\!\!\int\limits_0^\infty {dx\over \pi}
{\left(1+|\phi|^2\ln(1\!+\!{1\over x^2})\right) \ln{(1\!+\!{1\over
x^2})}\over \left(1\!+|\phi|^2\ln(1\!+\!{1\over
x^2})\right)^2\!\!+ x^2U^2\ln^2(1\!+\!{1\over x^2})},
\label{eq:10}
\end{equation}
The RHS of (\ref{eq:10}) is bounded from above with the value $1$
achieved at $\phi=0$ and $U=0$. Therefore we conclude that if $J$
is less than the critical value $J_c$ given by
\begin{equation}
{1\over J_c(U)}=\int\limits_0^\infty {dx\over \pi} {\ln{(1+{1\over
x^2})}\over 1+x^2U^2\ln^2\!(1+{1\over x^2})},
\end{equation}
then $\phi=0$ at the saddle point and the impurity is decoupled
from quasiparticles. The critical coupling is an increasing
function of $U$. In particular
\begin{eqnarray}
&&J_c(U)\approx 1+{1\over 3}U^2\quad \mbox{at}\;U\ll 1,\nonumber\\
&&J_c(U)\approx {U}\qquad\qquad \mbox{at}\; U\gg 1.
\label{eq:24}
\end{eqnarray}
As $J$ increases and becomes larger than $J_c$, the nontrivial
solution of (\ref{eq:10}) becomes relevant, since it defines the
minimum of the action as a function of $|\phi|^2$. Near the
critical point $J\approx J_c$ we have:
\begin{eqnarray}
&&|\phi|^2\approx {J - J_c\over 4 J_c^2\ln\!2 } \left(1+{U^2\over
2\ln 2}\right)\qquad\;\; \mbox{at}\; U\ll 1,\nonumber\\
&&|\phi|^2\approx {J - J_c(U)\over
J_c(U)}\qquad\qquad\qquad\quad\, \mbox{at}\; U\gg 1,
\end{eqnarray}
At the transition $T_K=0$ and according to (\ref{T_K}) its
asymptotics are:
\begin{equation}
\renewcommand{\arraystretch}{2.3}
\begin{array}{ll}
T_K(J, U)\approx \exp\left(-{J_c^2(U) \ln\!2\over 2 (J - J_c(U))^2
U}\right) & \mbox{at}\; U\ll 1,
\\
T_K(J, U)\approx \exp\left(-{J_c^2(U)\over 4(J-J_c(U))^2 \ln\!U
}\right) & \mbox{at}\; U\gg 1.
\end{array}
\end{equation}

These expressions show that close to the phase transition $T_K$ is
exponentially small and as long as $T_K<T\ll 1$, the system is in
the regime where the effective impurity energy $\epsilon$ is a
temperature dependent quantity (see (\ref{3u})):
\begin{eqnarray}
\epsilon&\approx& {1\over \beta} \left(1+{J- J_c\over J_c^2 2
\ln2} \ln\beta\right)^2 {(J-J_c)U \over J_c^2}\;\; \mbox{at}\;
U\ll 1,
\nonumber\\
\epsilon&\approx& {1\over\beta} \left(1+2 {J - J_c(U)\over
J_c(U)}\ln\beta\right)^2\nonumber\\&\times& {2(J - J_c(U))\ln U
\over J_c(U)+ 2(J-J_c(U))\ln U}\quad\qquad  \mbox{at}\;U\gg 1.
\label{eq:38}
\end{eqnarray}
The exponential behavior of the Kondo temperature and the
effective impurity energy has been predicted already
in~\cite{Fradkin1}. However, we note that this is the case only if
the particle-hole symmetry is broken. In the particle-hole
symmetric case $\epsilon$ and $T_K$ are identically equal to zero
for the whole range of the magnetic coupling constant. In the
regime, where $T\ll~T_K$, $\epsilon$ saturates at the value
proportional to $T_K$ (\ref{24}).

As we noted in the introduction, our results are quite different
from those derived in~\cite{Fradkin,Fradkin1}. We believe there
are several weak points in those treatments. In particular, the
action (2.28) in~\cite{Fradkin} clearly doesn't give the zero
saddle point for the effective impurity energy $\epsilon_f$ at the
particle-hole symmetric case $Q_f=N_c/2$ (we use notations
of~\cite{Fradkin} here). As a result the action in that form
doesn't predict the Curie law for the susceptibility even when the
impurity is free. On the other hand this symmetrization is
appropriate for finding the order parameter $\phi$ and the
critical coupling $g_c$ ($J_c$ in our notation). That is why we
find completely different expressions for the Kondo temperature as
compared to~\cite{Fradkin}, while the critical coupling $J_c$
remains the same. In their second paper~\cite{Fradkin1}, the
authors considered only the linear density of states (see (3.33)
therein) and treated the action more carefully. However the saddle
point equation (5.3) in ~\cite{Fradkin1} is still incorrect. It
gives the nonzero value of $\epsilon_f$ in the particle-hole
symmetric case (see e.g. (5.9) with $x=1/2$). It seems that there
was lost a contribution to (5.3) from the residue at
$|\epsilon|>|\Delta_0|$, and hence an implicit particle-hole
asymmetry was introduced.

Figure 1 illustrates the phase diagram in the $J- U$ plane. The
solid line corresponds to the phase transition, while the dash and
dot lines show the contours of the constant $T_K$. At fixed $T$
the latter lines separate high-temperature ($T_K<T$) and the
low-temperature ($T_K>T$) regions. Here we would like to emphasize
that the phase diagram with the dash line being a boundary is
reminiscent to that obtained by NRG~\cite{Ing2}. This probably
indicates that the large N-theory is not very reliable in the zero
temperature limit. It is also important to note that the curve
$J(U)$, corresponding to $T_K=$const is very insensitive to the
variations of the latter. Thus the dot and the dash  lines
corresponding to $T_K=10^{-4}$ and $T_K=10^{-5}$ almost coincide.

\begin{figure}
\label{fig1}
\includegraphics[angle=90, width=7.2cm]{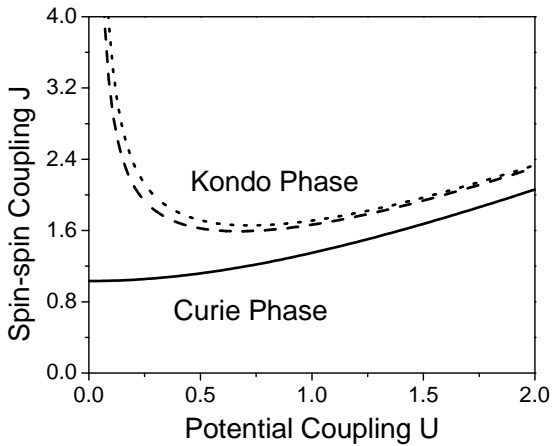}
\caption{Phase diagram in the $J\!-\!U$ plane. The solid line
($J=J_c$) corresponds to the Kondo transition at zero temperature.
At this coupling the order parameter $\phi$ first becomes
non-zero. The dot and the dash line are the contours of constant
Kondo temperature $T_K(J,U)=10^{-4}$ and $T_K=10^{-5}$
respectively. The critical spin-spin coupling $J_c$ remains finite
at $U=0$, however in the particle-hole symmetric case there is no
real transition to the Kondo phase since at any finite temperature
the crossover value of $J$ (dash and dot lines) diverges as $U\to
0$. For $r<1$ the phase diagram is similar to this, while for
$r>1$ the zero temperature boundary between the Kondo and Curie
phases is analogous to the dash line. The solid line in this case
doesn't represent any real phase transition.}
\end{figure}

\subsection{Generalization to the nonlinear density of states}

Before we proceed further with D-wave superconductors, let us
generalize obtained in the large N limit expressions to the case
of vanishing density of states with an arbitrary exponent:
\begin{equation}
\renewcommand{\arraystretch}{1.5}
\rho(\epsilon)=\left\{
\begin{array}{cl}
{r+1\over 2} |\epsilon|^r & |\epsilon|<1\\ 0 & |\epsilon|\geq 1
\end{array}\right.
\end{equation}
I.e. the action is still given by the equation (\ref{main}) but
the unperturbed Green function $G_0^0(\omega_n)$ is now as
follows:
\begin{eqnarray}
G_0^0(\omega_n)\!=\! i\!\!\int\limits_{-1}^1
{\omega_n\rho(\epsilon) \over \omega_n^2\!\!+\!\epsilon^2}
d\epsilon=i {r\!+\!1\over 2} {{\cal F} \left(1,\!{1+r\over
2},\!{3+r\over 2},\! -{1\over \omega_n^2}\right)\over
\omega_n},\phantom{x}
\end{eqnarray}
where ${\cal F}$ is the hypergeometric function. High and low
frequency asymptotics of $G_0^0$ are
\begin{equation}
G_0^0(\omega_n)\approx\left\{
\renewcommand{\arraystretch}{1.8}
\begin{array}{cl}
i\omega_n^r {r+1\over 2} {\pi\over \cos {\pi r\over 2}}-i\omega_n
{r+1\over r-1} & \omega_n\ll 1\\ {i\omega_n^{-1}} & \omega_n\gg 1.
\end{array}
\right. \label{eq:6}
\end{equation}
For $r< 1/2$ the problem was studied in detail
in~\cite{Ing1,Withoff}. Both the consequent analysis and numerical
results~\cite{Ing2} show that the effects of the particle-hole
asymmetry become crucial for $r>1/2$. Here only this situation
will be considered. Doing the same steps as before we find the
critical $J$, where $|\phi|$ first becomes nonzero, is
\begin{eqnarray}
&&J_c\sim {r(r+1)\over 2} \qquad\qquad\quad\; \mbox{at}\;\;U\to
0\nonumber\\ &&J_c\sim U\, {\rm max} \left({2r\over r+1},1\right)
\quad \mbox{at}\;\;U\to\infty.
\label{eq:5}
\end{eqnarray}
In the case of the small potential scattering $U$, the order
parameter near the phase transition is:
\begin{equation}
|\phi|^2\!\sim {J-J_c\over J_c^2}{4r-2\over (r+1)^2}\left(
\psi\left({r+1\over 2}\right)-\psi\left({1\over
2}\right)\right)^{-1}\!\!, \label{eq:4}
\end{equation}
where $\psi$ is the digamma function. Note that regardless of $r$,
$|\phi|^2$ is proportional to $J-J_c$. However, when $\epsilon$ is
considered, it is necessary to distinguish between $r<1$ and
$r>1$. In the former case one can define the Kondo temperature:
\begin{equation}
T_K\approx \left[{2\cos{\pi r\over 2}\over \pi^r
(1-r)}+{4^r-1\over 4^{r-1}}{\cos{\pi r\over 2}\zeta(2r)\over
\pi^{1+r}|\phi|^4 F_r(U)}\right]^{-{1\over 1-r}}, \label{eq:13k}
\end{equation}
where $\zeta$ is the Riemann zeta function,
\begin{eqnarray}
&&\hspace{-1cm}F_r(U)\!\sim\! {U\over 2}{(r\!+\!1)^2\over 2r-1}
\left(\psi\!\left({1\!+\!r\over 2}\right)\!-\!\psi\!\left({r\over
2}\right)\right)\; \mbox{for}\; U\ll 1\nonumber
\\
&&\hspace{-1cm}F_r(U)\!\sim\! {U^{1-r\over r}-1\over
\left(\cos{\pi r\over 2}\right)^{1\over r}} {\pi^{1\over
r}(1+r)^{1\over r} \over r\,2^{1+r\over r}\sin{\pi \over
2r}}\qquad\qquad \mbox{for}\; U\gg 1. \label{eq:12}
\end{eqnarray}
For $r<1$ near the phase transition we have:
\begin{equation}
T_K\propto (J-J_c)^{2\over {1-r}}
\end{equation}
Above the Kondo temperature $\epsilon$ increases with $T$, while
for $T<T_K$ it saturates at the value:
\begin{equation}
\epsilon\sim |\phi|^2 {\pi^{1+r}(1+r)\over 2\cos{\pi r\over
2}}T_K^{r}\left(1-{2\cos{\pi r\over 2}
\over\pi^r(1+r)}T_K^{1-r}\right).
\end{equation}
It is easy to check that the expressions for $r=1$ obtained
earlier are consistent with these formulas.

The picture becomes quite different for $r>1$. If the order
parameter is sufficiently small, $\epsilon$ can be found from:
\begin{equation}
\tanh \left({\beta \epsilon\over 1+{r+1\over
r-1}|\phi|^2}\right)\!\! =2|\phi|^2\! \left(1\!+\! {r\!+\!1\over
r\!-\!1}|\phi|^2\right)\! F_r(U). \label{eq:16a}
\end{equation}
Provided the RHS of (\ref{eq:16a}) is less than 1, this equation
has a solution with $\epsilon\propto T$, i.e. vanishing as $T$
goes to zero. On the other hand if the RHS becomes greater than 1,
this equation has no solution and at zero temperature at this
point there is a phase transition to the Kondo phase. Thus for $r$
slightly greater than $1$ and $U\ll 1$ the new critical coupling
$\tilde J_c$ is
\begin{equation}
\tilde J_c\sim J_c + 0.35 \sqrt{r-1\over U}.
\label{hru}
\end{equation}
For $U\gg 1$ we have $\tilde J_c\approx J_c$. The diverging
behavior of $\tilde J_c$ at $U\to 0$ is in fact very similar to
that of the crossover spin-spin coupling for $r=1$ at finite
temperature (see dash line on fig. 1). However, for $r>1$ this
divergence exists even in the zero temperature limit.

At this point we get another evidence that the large-N theory
might be not reliable at small temperatures for $r\leq 1$. Namely,
the NRG analysis shows that $r=1$ is not a special point in the
phase diagram~\cite{Vojta1}. On the other hand, the saddle point
results predict that the curve similar to the dash (not solid)
line in figure 1 represents the actual transition to the Kondo
phase for $r>1$. At the same time at finite temperature the mean
field phase diagram is qualitatively similar to the NRG
result~\cite{Ing1, Vojta1}. And moreover, for $r$ close to 1 the
contours $T_K(J,U)=T$ are very insensitive to the value of $T$
(see figure 1).

\section{Magnetic susceptibility and quasiparticle density of states.}

We start this section from calculating a local magnetic
susceptibility at zero magnetic field. In the saddle point
approximation it is given by:
\begin{equation}
\chi_{loc}=-{1\over 4\beta}{\partial^2 S_{eff}(h)\over \partial
h^2}=- {1\over 4\beta}{\partial^2 S_{eff}(\epsilon) \over
\partial \epsilon^2}.
\end{equation}
Using (\ref{seffeps}) we immediately get:
\begin{equation}
\chi_{loc}=\sum_{\omega_n}{N\over 4\beta \left(\omega_n-i\epsilon-
i |\phi|^2 G_\uparrow(\omega_n)\right)^2}. \label{eq:14}
\end{equation}
This expression becomes particularly simple if there is no
potential scattering and $\epsilon=0$ (we always assume that the
temperature is small compared to the cutoff, i.e. $\beta\gg 1$):
\begin{equation}
\chi_{loc}\approx{N \beta\over 16}\left(1+2|\phi|^2
\log{\beta\over \pi}\right)^{-2}. \label{eq:13}
\end{equation}
It is not surprising that for $\phi=0$ in the physical case $N=2$,
(\ref{eq:13}) gives one half of the usual Curie constant. This
feature is an artifact of the large N techniques applied here. For
$r<1$ instead of (\ref{eq:13}) we have
\begin{eqnarray}
&&\chi_{loc}\approx{N \beta\over 16} \quad 1\ll
\beta\ll \beta_0\equiv\pi \left[{2\cos{\pi r\over 2}\over \pi (r+1)
\phi^2}\right]^{1\over 1-r }\label{eq:13d}\\ &&\chi_{loc}\approx
{2 N\beta^{2r-1}\cos^2{\pi r\over 2}\over \phi^4 \pi^{2(r+1)}
(r+1)^2} {4^r -1\over 4^r}\zeta(2r) \;\;
\beta\gg \beta_0.
\label{eq:13a}
\end{eqnarray}
\begin{figure}
\label{fig2}
\includegraphics[angle=90, width=7cm]{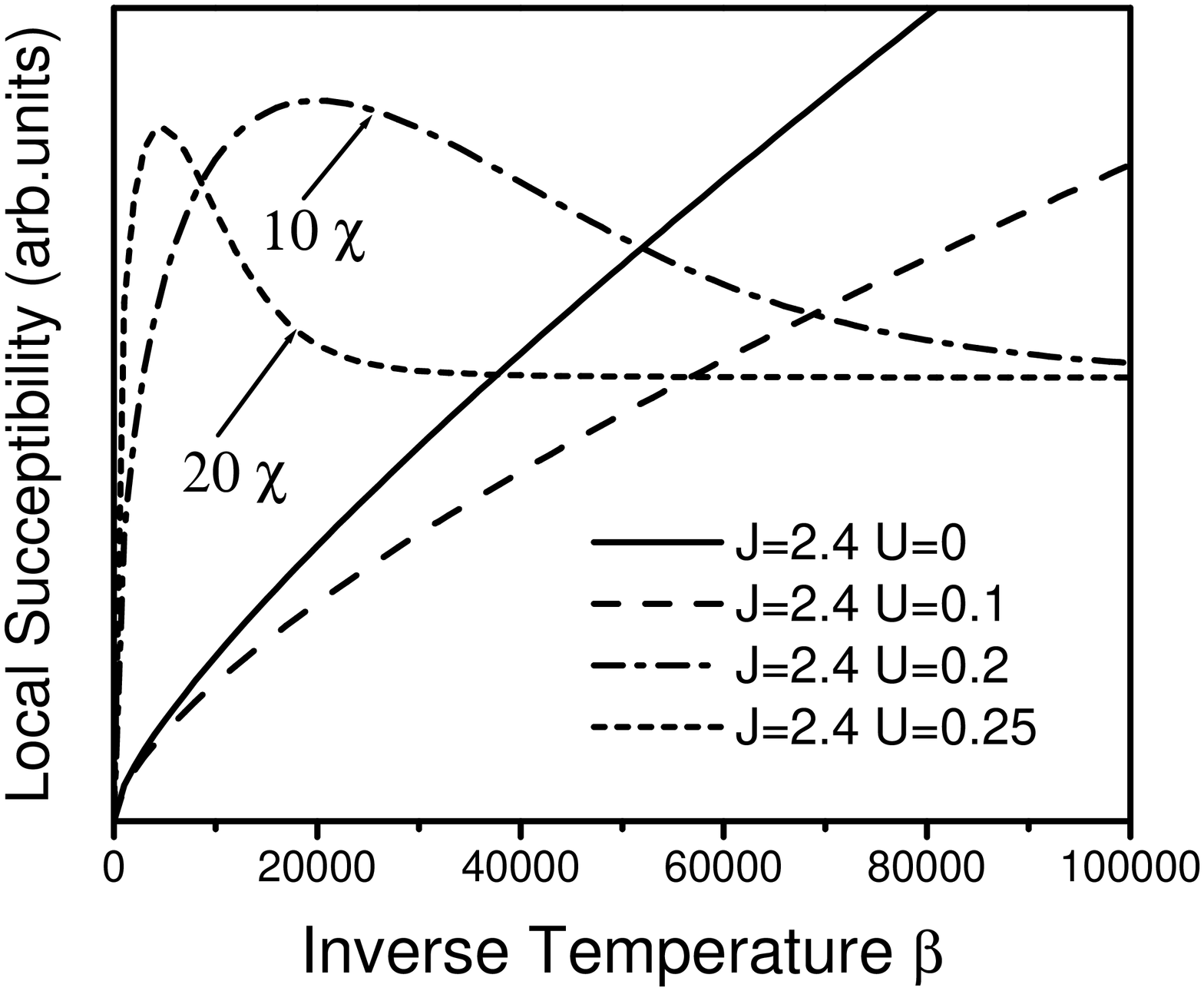}
\includegraphics[angle=90, width=7cm]{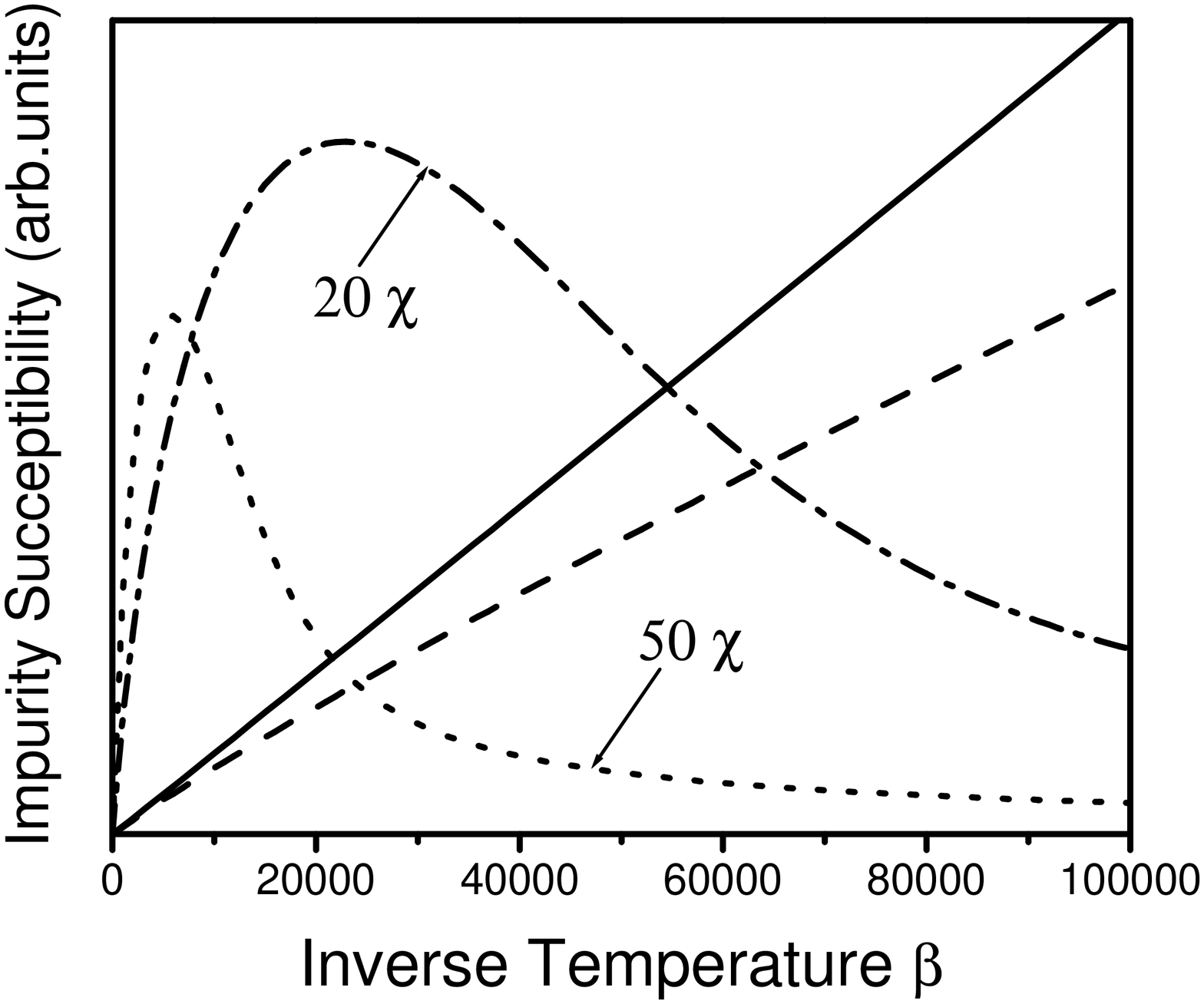}
\caption{Temperature dependence of the local susceptibility (top)
and the impurity susceptibility (bottom) in the Kondo phase for
different potential couplings. For $U=0$ and $U=0.1$ (see Figure
1) the Kondo temperature is much smaller than $T=\beta^{-1}$ for
all plotted values of $\beta$. So both susceptibilities increase
with $\beta$. Note that $\chi_{loc}$ is clearly a sublinear
function of inverse temperature even in the particle-hole
symmetric case.  For $U=0.2$ and $U=0.25$ the Kondo temperature
becomes large enough and both susceptibilities have a maximum at
$\beta T_K\approx 1$. The local susceptibility saturates at
$\beta\to\infty$, while $\chi_{imp}\to 0$ (see Eq. \ref{eq:40}).}
\end{figure}
On the other hand for $r>1$ the local susceptibility in the
particle-hole symmetric case is completely described by the Curie
law.

The formulas above are valid as long as $\epsilon\ll T$. For
$r\leq 1$ this is equivalent to $T\gg T_K$. Below the Kondo
temperature $\chi_{loc}$ saturates at:
\begin{eqnarray}
\chi_{loc}&\approx& {N |\phi|^2\over 4 T_K
(1-2|\phi|^2\ln(T_K))^3}\qquad\;\;\,\mbox{for}\;r=1,\\
\chi_{loc}&\approx& {N\over 4 |\phi|^4
T_K^{2r-1}}{2(1-r)\cos^{2}{\pi r\over 2}\over
r^2\pi^{2r-1}(r+1)^2}\quad\mbox{for}\;r<1.
\end{eqnarray}

The temperature dependence of $\chi_{loc}$ for $r=1$ is plotted on
the top graph on Figure 2. This function is non-monotonic with the
maximum occurring at $T\approx T_K$.

Using equation (\ref{eq:14}), it is not hard to calculate a
correction to $\chi_{loc}(T=0)$ for $T\ll T_K$:
\begin{equation}
\chi_{loc}(T)-\chi_{loc}(0)\approx {9\zeta(3)|\phi|^2 T^3\over
\pi\epsilon^4}\left(1+2|\phi|^2 \ln {1\over T}\right),
\label{eq:15}
\end{equation}
showing that $\chi_{loc}$ decreases at $T\to 0$ for $T<T_K$.

The local susceptibility is the response function to the magnetic
field coupled only to the impurity. We will also consider here the
impurity susceptibility which additionally includes the
contribution from the nearby electron cloud .

Assuming that g-factors of band and impurity electrons are the
same we see that the only effect of the magnetic field $h$ on the
action (\ref{main}) is the shift of the imaginary frequency
$\omega_n\to \omega_n+ih$ for the spin up states and $\omega_n\to
\omega_n-ih$ for the spin down states. Therefore, the impurity
susceptibility at zero field is:
\begin{eqnarray}
\chi_{imp}&=&{N\over 4\beta} \sum_{n}
{\left(1-i|\phi|^2{\partial\over
\partial \omega_n} G_\uparrow(\omega_n)\right)^2\over \left(\omega_n+i\epsilon- i|\phi|^2
G_\uparrow(\omega_n)\right)^2}\nonumber\\&-&{N\over 4\beta}
\sum_{n}{|\phi|^2 {\partial^2\over \partial \omega_n^2}
G_\uparrow(\omega_n)\over i\omega_n-\epsilon+|\phi|^2
G_\uparrow(\omega_n)}. \label{eq:16}
\end{eqnarray}
%
If $\epsilon=0$, expression above reduces to:
\begin{equation}
\chi_{imp}\approx {N\beta\over 16}\left(1-{2|\phi|^2\over
1+2|\phi|^2\ln{\beta}}\right).
\end{equation}
Note that $\chi_{imp}$ asymptotically approaches Curie law at
$T\to 0$, contrary to (\ref{eq:13}) where $T\chi_{loc}\to 0$. This
peculiarity was already predicted in~\cite{Chen}, however in the
later work~\cite{Ing2}, $\chi_{loc}$ was found to be proportional
to the inverse temperature without any logarithmic corrections.
The controversy between the mean-field and NRG results can not be
resolved considering the next terms in the $1/N$ expansion near
the saddle point. In fact, it is easy to show, that for $U=0$ at
each order of the perturbation series for $\chi_{loc}$, a
multiplier $(1+2|\phi|^2\ln\beta)^2$ is generated in the
denominator. Therefore the sublinear dependence of $\chi_{loc}(T)$
is generic for the large $N$ approach. Probably the reason for the
discrepancy between this work and~\cite{Ing2,Vojta1} is that the
large $N$ expansion is asymptotic in the particle-hole symmetric
case and for $r<1$, and it diverges as temperature goes to zero.

For $r>1$ near $J=J_c$, $\chi_{imp}=\chi_{loc}=N\beta/16$. While
for $r<1$ there is a crossover between
\begin{equation}
\chi_{imp}\approx {N\beta\over 16}
\end{equation}
and
\begin{equation}
\chi_{imp}\approx {Nr\beta\over 16},
\end{equation}
$T\beta_0\gg 1$ and $T\beta_0\ll 1$ respectively, $\beta_0$ is
defined in (\ref{eq:13d}). So the impurity is partially screened.
We note again that these results are valid only for $r>1/2$,
otherwise there is a transition to Kondo phase even in the
particle-hole symmetric case.

Below the Kondo temperature $\chi_{imp}$ goes to zero in agrement
with~\cite{Fradkin1} . For example at $r=1$:
\begin{equation}
\chi_{imp}\approx {T |\phi|^2 N \ln 2 \over T_K^2
\left(1-2|\phi|^2\ln T\right)}. \label{eq:40}
\end{equation}

Impurity susceptibility as a function of temperature is plotted on
the bottom graph on Figure 2.

Another manifestation of the coupling between quasiparticles and
impurity is the change of the electron density of states, which is
given by the imaginary part of the effective Green function:
\begin{equation}
\rho (\Omega,{\bf r})=\Im\, {\rm Tr}{1+\tau_z\over 2} \tilde
G({\bf r},{\bf r},\Omega),
\label{eq:18}
\end{equation}
where $\Omega$ is the real frequency with an infinitesimal
positive imaginary part,
\begin{widetext}
\begin{equation}
\tilde G({\bf r},{\bf r}^\prime,\Omega)=G({\bf r},{\bf
r}^\prime,\Omega)+G({\bf r},{\bf r}_0,\Omega)\, {\cal
T}_K(\Omega)\, G({\bf r}_0,{\bf r}^\prime,\Omega). \label{a2}
\end{equation}
The scattering matrix ${\cal T}_K(\Omega)$, corresponding to the
Kondo contribution is equal to:
\begin{equation}
{\cal T}_K(\Omega)={|\phi|^2 \over -\Omega+\epsilon\tau_z-|\phi|^2
\tau_z G({\bf r}_0,{\bf r}_0,\Omega)\tau_z}, \label{a3}
\end{equation}
note that a similar form for ${\cal T}$ was found in~\cite{Zhang}
for the case of an Anderson impurity below the Kondo temperature.
Using (\ref{2}), (\ref{eq:18}) can be simplified further:
\begin{eqnarray}
\rho (\Omega,{\bf r})= \Im\; G_+^0({\bf r},\Omega){\cal
T}_+(\Omega)G_+^0(-{\bf r},\Omega)+G_x^0({\bf r},\Omega){\cal
T}_-(\Omega)G_x^0(-{\bf r},\Omega).
\label{eq:18c}
\end{eqnarray}
We adopted the notation:
\begin{equation}
G_+^0({\bf r},\Omega)={\rm Tr}\, {1+\tau_z\over 2}\, G_0({\bf
r},\Omega),\quad G_x^0({\bf r},\Omega)={\rm Tr}\, {\tau_x\over
2}\, G_0({\bf r},\Omega),
\end{equation}
\begin{equation}
{\cal T}_\pm (\Omega)=-{1\over 1\pm U G_0^0(\Omega)}+
\left({1\over 1\pm U G_0^0(\Omega)}\right)^2 {1\over -\Omega\pm
\epsilon-|\phi|^2 {G_0^0(\Omega)\over 1\pm U G_0^0(\Omega)}}.
\end{equation}
\end{widetext}

The spatial dependence of DOS coincides with that obtained earlier
for the case of the pure classical scattering~\cite{Bal}. We note
that at small frequencies ($\Omega\ll \Delta$), there is a maximum
at the nearest neighbors of the impurity and at $r\to \infty$ the
asymptotic behavior of the DOS is $r^{-2}$. Also it is easy to
show that if the resonance frequency is much smaller than the
superconducting gap, the spatially integrated DOS remains
particle-hole symmetric. It is determined by the momenta in the
vicinity of the nodes and therefore this result is valid
regardless the microscopical details of the model such as chemical
potential, existence of second-neighbor hopping matrix elements,
etc.

Clearly the ${\cal T}$-matrix for the Kondo scattering channel has
a pole at the frequency $\Omega\approx T_K$. If $T_K\ll \Delta$,
i.e. $J\approx J_c$, the resonant levels corresponding to the
poles in ${\cal T}_+$ and ${\cal T}_-$ become very narrow and they
have roughly Lorentzian shape. In the opposite limit $\Delta\ll
T_K$, the resonance corresponding to ${\cal T_-}$ disappears and
(\ref{eq:18c}) results in the Fano lineshape in agreement
with~\cite{Zawadowski}.

We would like to emphasize that the scattering on a strong
classical magnet gives two identical peaks at positive and
negative frequencies and additional weak potential coupling
results in the weak splitting of those~\cite{Bal}. In the Kondo
case, even at small $U$ the spatial structure of the resonances at
positive and negative frequency is highly asymmetric.

\section{Non-magnetic impurity in a D-wave superconductor}

In this section we focus on the situation, when the spin is not
located on the impurity site, but rather it is a staggered moment
distributed nearby. The main motivation to this investigation
comes from the NMR
experiments~\cite{Alloul,Farlane,Mahajan,Bobroff,Mendels,Julien}
with non-magnetic Zn or Li impurities in a BSSCO. The effective
strength of the induced moments rapidly decreases with the
distance~\cite{Julien}, therefore it is reasonable to assume that
the spins are sitting only on the first nearest neighbors.

In this way the new interacting part of the Hamiltonian becomes:
\begin{widetext}
\begin{equation}
H_{int}=U\psi^\dagger_i ({\bf r}_0)\tau_z^{ij}\psi_j ({\bf
r}_0)-{J\over 2}\sum_{s\in O} (\psi^\dagger_\uparrow ({\bf r}_s)
F_\uparrow-F^\dagger_\downarrow \psi_\downarrow({\bf r}_s))
(F_\uparrow^\dagger\psi_\uparrow ({\bf
r}_s)-\psi^\dagger_\downarrow({\bf r}_s)F_\downarrow),
\end{equation}
where $O$ denotes the subset of the sites closest to the ${\bf
r}_0$. Contrary to the spin-spin coupling which is strongest on
the nearest neighbors of the impurity, the potential scattering is
dominated by the local term, therefore its form is unchanged as
compared to (\ref{hint1}). Repeating the same steps as before and
using the static approximation we obtain:
\begin{equation}
S=-{\rm Tr}\, \ln\left({\partial\over\partial\tau}
+\epsilon\tau_z-\sum_{s,s^\prime\in O}\phi_s\,G\left({\bf
r}_s,{\bf r}_{s^\prime},{\partial\over\partial\tau}
\right)\phi_{s^\prime}\right)+{2\beta\over J}\sum_{s\in O}
\phi_s^2, \label{maine}
\end{equation}
\end{widetext}
This action can be also evaluated in the saddle point
approximation. The important difference with the local spin case
is that there are four nontrivial saddle points corresponding to
$S$
($\phi_{10}=\phi_{01}=\phi_{\overline{1},0}=\phi_{0,\overline{1}}$),
$D$
($\phi_{10}=-\phi_{01}=\phi_{\overline{1},0}=-\phi_{0,\overline{1}}$),
and $P$ ($\phi_{10}=-\phi_{\overline{1},0}\,$;
$\phi_{0,1}=\phi_{0,\overline{1}}=0$) symmetry of the order
parameter, the latter case being twice degenerate. On the Figure 3
we show the dependence of the critical coupling $J_c$ on the
potential scattering for the order parameter of the $S$ and $D$
symmetry ($P$ case is always in between) and for the local spin.
Clearly, for the whole range of $U$, $D$ order has a lower $J_c$
and as a result a smaller action. Also, contrary to the local spin
case, $J_c$ decreases with $U$. This phenomenon is readily
understood since the potential scattering increases the
quasiparticle density of states on the nearest
neighbors~\cite{Bal}, where the magnetic moments are located, and
hence enhances the effective spin-spin coupling. It is interesting
to note, that the effective DOS of the "bath" electrons for the
$D$-wave saddle point in the absence of potential scattering term
is proportional to $\omega^3$ and not $\omega$ due to Green
functions cancellation near the nodes~\cite{Vojta2}. However this
feature is not important for our consequent analysis.

\begin{figure}
\label{fig3}
\includegraphics[angle=90, width=7.5cm]{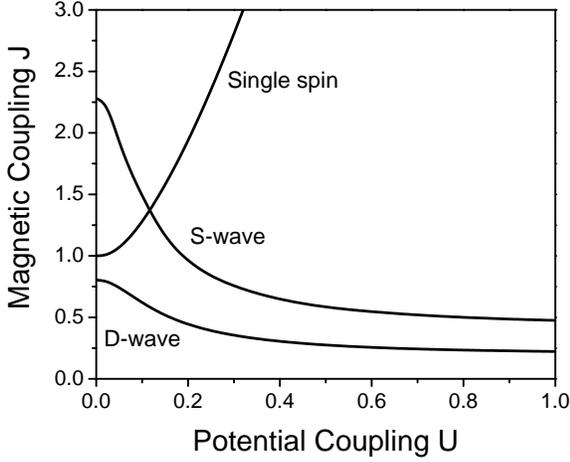}
\caption{Critical spin-spin coupling as a function of potential
scattering for the non-local spin case. The D-wave configuration
of the order parameter $\phi$ always has a lower critical coupling
$J_c$ and the smallest action. Contrary to the local spin case the
Kondo transition occurs at lower value of $J$ with increasing
$U$.}
\end{figure}

Both local and impurity magnetic susceptibility practically don't
change as compared to the local spin case, since they are
sensitive to the value of the impurity spin and not to its
distribution. The situation is quite different for the density of
states . In particular, for the spin scattering part and $D$-wave
order parameter (\ref{eq:18c}) generalizes to
\begin{eqnarray}
\rho (\Omega,{\bf r})&=& \Im\; G_+^d({\bf r},\Omega){\cal
T}_+(\Omega)G_+^d(-{\bf r},\Omega)\nonumber\\&+&G_{x}^d({\bf
r},\Omega){\cal T}_-(\Omega)G_x^d(-{\bf r},\Omega),
\label{eq:18d}
\end{eqnarray}
where:
\begin{equation}
{\cal T}_\pm(\Omega)={-1\over \Omega\mp \epsilon\!+\!|\phi|^2
\left(G_0^d(\Omega)\!+\! {G_x^{d^2}({\bf r}_0,\Omega) U^2
G_0^0(\Omega)\over 1- U^2G_0^{0^2}(\Omega)}\right)},
\end{equation}
\begin{eqnarray}
&&G_{+,x}^d({\bf r},\Omega)=\sum_{s\in O} (s_x^2-s_y^2)\,
G_{+,x}({\bf r},{\bf r}_s,\Omega),\nonumber\\  &&
G_0^d(\Omega)=\sum_{s,s^\prime\in O}
(s_x^2-s_y^2)(s_x^{\,\prime\,2}-s_y^{\,\prime\,2})\,G_0^0({\bf
r}_s-{\bf r}_{s^\prime},\Omega),\nonumber
\end{eqnarray}
with $s_i=0,\pm 1$. The resonance frequency is approximately equal
to the Kondo temperature as in the local spin case. If the
potential scattering is not too large $UG_0^0(T_K)\ll 1$ then the
spatial dependence of the DOS is determined by the interference
from the scattering on the four nearest neighbors to the impurity
sites. In particular, for the $D$ ($S$) order parameter, the
resonance corresponding to ${\cal T}_-$ ($\cal T_+$) matrix gives
the largest DOS on the impurity site ${\bf r}={\bf r}_0$, local
maximum on the second neighbors and very small signal on the first
neighbors~\cite{polk}. This behavior is in the excellent agreement
with the experiments, where the spatial distribution of DOS near
the Zn impurity was directly observed~\cite{Davis}. Also at $r\to
\infty$ the density of states vanishes as $r^{-4}$, i.e. much
faster than for the the local spin case. The spatially integrated
DOS is generically particle-hole asymmetric. In particular it can
be shown that if $UG_0^0(T_K)\ll 1$ and $T_K\ll \Delta$ then at
the resonance frequency the integrated DOS is:
\begin{eqnarray}
\rho_{+}(\Omega)&\approx& {4\over \Delta_0^2}\int {\varepsilon_k^2
\Delta_k^2\over \left(\varepsilon_k^2+\Delta_k^2\right)^2} {d^2
k\over (2\pi)^2}\Im {\cal T}_+(\Omega)
\label{101}
\\
\rho_{-}(-\Omega)&\approx& {4\over \Delta_0^2}\int {\Delta_k^4
\over \left(\varepsilon_k^2+\Delta_k^2\right)^2} {d^2 k\over
(2\pi)^2}\Im {\cal T}_-(-\Omega)
\label{102}
\end{eqnarray}
Simple power counting shows that (\ref{101}) and (\ref{102})
converge near the nodes. Therefore the ratio of the spatially
integrated DOS is nonuniversal and while for particle-hole
symmetric spectrum $\rho_-/\rho_+\sim 2.5$, it decreases with the
doping crossing 1 at some moment. For comparison we will give the
results for the local spin:
\begin{eqnarray}
\tilde\rho_{+}(\Omega)&\approx&\!\!\!\!\!\!\!\!\!\!
\int\limits_{\varepsilon_k^2+\Delta_k^2>\Omega^2} {\varepsilon_k^2
\over \left(\varepsilon_k^2+\Delta_k^2\right)^2} {d^2 k\over
(2\pi)^2}\Im {\cal T}_+(\Omega)
\label{103}
\\
\tilde\rho_{-}(-\Omega)&\approx&\!\!\!\!\!\!\!\!\!\!
\int\limits_{\varepsilon_k^2+\Delta_k^2>\Omega^2} {\Delta_k^2
\over \left(\varepsilon_k^2+\Delta_k^2\right)^2} {d^2 k\over
(2\pi)^2}\Im {\cal T}_-(-\Omega)
\label{104}
\end{eqnarray}
This expressions logarithmically diverge when $\varepsilon$ and
$\Delta$ become small, therefore it is necessary to put a low
energy cutoff of the order of frequency. The integrals mainly sit
on the nodes and $\rho_+\approx \rho_-$ if $\Omega$ is small.
Physically this differences are related to the slow ($1/r^2$)
dependence of the DOS on the distance for a single spin, while for
the nonlocal scattering the DOS rapidly decreases away from the
impurity and the main contribution to the spatial integrals comes
from the short distances or high momenta.

Let us say a few words about strong potential scattering case
$U\gg 1$. As we see from the Figure 3, the critical Kondo coupling
$J_c$ goes to zero, in fact as $1/U$ up to logarithmic
corrections. Therefore for the most values of $J$ the system is in
the Kondo phase. If $J$ is not too large, then the order parameter
$|\phi|$ and the Kondo temperature are small enough so that the
condition $UG_0^0(T_K)\ll 1$ fulfills. As a result the spatial
structure of the DOS remains the same as if there is no potential
scattering. This contradicts the naive idea of the "hard-wall"
impurity, which prohibits large DOS at ${\bf r}={\bf r}_0$. And
the reason, why this argument is not correct is that the energy
scale $T_K$ is so small that the potential scattering is
irrelevant. If both conditions $U\gg 1$ and $J\gg J_c$ are
fulfilled then $UG_0^0(T_K)$ might be greater than 1. In this case
the Kondo scattering occurs on the almost frozen density of states
created by the potential scatterer and apart from small variations
the spatial structures of the Kondo and potential resonances
coincide. But we emphasize again that in order to have this
situation both $U$ and $J$ should be very large.

\section{Conclusions}

We have studied the screening of a magnetic impurity in $D$-wave
superconductors. It is found that there is a Kondo transition,
which occurs only if the particle-hole symmetry is distorted, e.g.
by the potential scattering. The Kondo temperature is
exponentially small near the critical coupling in agreement
with~\cite{Fradkin1}. However, at finite but small temperature the
crossover from the regime $T\gg T_K$, where the impurity spin is
not screened to $T\ll T_K$ occurs at qualitatively different
values of the magnetic coupling $J$ (see (fig. 1)), in particular
$J\to\infty$ if the potential coupling $U\to 0$. In the
particle-hole symmetric phase it was found that the impurity
susceptibility behaves in a Curie-like fashion: $T\chi_{imp}\to
const$ as $T\to zero$, while the local susceptibility has
multiplicative logarithmic corrections, so that: $T\chi_{loc}\to
0$. This result, which holds in any order of $1/N$ expansion,
agrees with~\cite{Chen}, however it contradicts later NRG
investigations~\cite{Ing2,Vojta1}. It is possible that the
logarithmic corrections are beyond the accuracy of the large $N$
expansion in the particle-hole symmetric case. On the other hand,
the resulting phase diagram at finite temperature is qualitatively
similar to that, obtained by NRG~\cite{Ing2,Vojta1} and dynamic
multichannel~\cite{Vojta} methods. Moreover the contours of
constant $T_K$ change very slowly (logarithmically) with $T_K$
(see fig. 1). Therefore, we expect that the conclusions about the
Kondo phase are reliable.

Also we examined the situations with $1/2<r<1$ and $r>1$. The
former appears to be analogous to the case $r=1$ with the only
difference that exponential (logarithmical) behavior is
substituted by the suitable power law. For $r>1$ the picture is
different in a sense that the phase diagram is qualitatively
consistent with NRG results even in the zero temperature limit.

Finally we studied the behavior of a non-magnetic impurity in a
D-wave superconductor, which creates a staggered magnetization on
the nearest copper neighbors. It was found that the saddle point
corresponding to the true minimum of the action always has a
$D$-symmetry. The resulting spatial structure of the quasiparticle
density of states near the impurity for this saddle point is
consistent with experimental data. It was also shown that unlike
the single spin case the critical magnetic coupling corresponding
to the Kondo transition decreases as $U$ increases .

\acknowledgments

The author is grateful to W.A. Atkinson and H. Alloul for useful
discussions and especially to S. Sachdev and M. Vojta for many
very important comments and suggestions.

\end{document}